# An efficient and low-cost method to create high-density nitrogen-vacancy centers in CVD diamond for sensing applications.

Prem Bahadur Karki,[1] Rupak Timalsina,[2] Mohammadjavad Dowran,[2] Ayodimeji E. Aregbesola,[1] Abdelghani Laraoui,[2,3*] and Kapildeb Ambal[1†]

[1]Department of Mathematics, Statistics, and Physics, Wichita State University, 1845 Fairmount St. Wichita, KS 67260, United States of America
[2]Department of Mechanical & Materials Engineering, University of Nebraska-Lincoln, 900 N 16th St. W342 NH. Lincoln, NE 68588, United States of America
[3]Department of Physics and Astronomy and the Nebraska Center for Materials and Nanoscience, University of Nebraska-Lincoln, 855 N 16th St, Lincoln, Nebraska 68588, USA
[*†]Author to whom correspondence should be addressed, email:
*alaraoui2@unl.edu, †Kapildeb.ambal@wichita.edu

## I. Abstract

The negatively charged Nitrogen-Vacancy (NV⁻) center in diamond is one of the most versatile and robust quantum sensors suitable for quantum technologies, including magnetic field and temperature sensors. For precision sensing applications, densely packed NV⁻ centers within a small volume are preferable due to benefiting from $1/\sqrt{N}$ sensitivity enhancement ($N$ is the number of sensing NV centers) and efficient excitation of NV centers. However, methods for quickly and efficiently forming high concentrations of NV⁻ centers are in development stage. We report an efficient, low-cost method for creating high-density NV⁻ centers production from a relatively low nitrogen concentration based on high-energy photons from Ar⁺ plasma. This study was done on type-IIa, single crystal, CVD-grown diamond substrates with an as-grown nitrogen concentration of 1 ppm. We estimate an NV⁻ density of ~ 0.57 ppm (57%) distributed homogeneously over 200 μm deep from the diamond surface facing the plasma source based on optically detected magnetic resonance and fluorescence confocal microscopy measurements. The created NV⁻s have a spin-lattice relaxation time ($T_1$) of 5 ms and a spin-spin coherence time ($T_2$) of 4 μs. We measure a DC magnetic field sensitivity of ~ 104 nT Hz$^{-1/2}$, an AC magnetic field sensitivity of ~ 0.12 pT Hz$^{-1/2}$, and demonstrate real-time magnetic field sensing at a rate over 10 mT s$^{-1}$ using an active sample volume of 0.2 μm³.

## II. Introduction

The negatively charged nitrogen-vacancy (NV⁻) centers are one of the leading solid state-based quantum platforms [1], enabling diverse quantum applications, including spin-qubit for quantum information processing [2–5], quantum memory [6–8], and quantum sensing [9] for many physical quantities including magnetic fields [10–17], electric fields [18–21], and temperature [22–24]. For potential commercial sensing applications where nanoscale spatial resolution is not required, ensemble NV⁻s are preferred. The key requirements for high-sensitive ensemble-based sensing applications are high NV⁻ concentrations within small volumes while preserving long spin coherence time. The high concentrations are favored for precision magnetometry due to improved signal-to-noise and sensitivity from $1/\sqrt{N}$ enhancement [9,25]. For example, the sensitivity to detect a constant magnetic field with NV⁻ centers scales with $1/\sqrt{N_{NV}}$ where $N_{NV}$ is the number of NV⁻ centers from which the fluorescence signal is collected [26]. The densely packed or small sample volume is necessary to efficiently excite NV$^-$ centers using on-chip laser and microwave excitation [27].

Several existing techniques are available for creating high-density NV$^-$ centers > 10 ppm (1 ppm = $1.76 \times 10^{17}$ cm$^{-3}$ in diamond) [28]. The most popular ones are; (*1*) implanting high doses (~100 ppm) of nitrogen (N) with subsequent annealing [29–31] and (2) electron irradiation or implantation of He$^+$ ions on a diamond substrate containing high-concentration of nitrogen (N) impurities (~100 ppm) followed by subsequent annealing [32,33]. Nevertheless, the final concentration of NV$^-$s still depends on many factors like annealing temperature, cleaning procedures, substrate characteristics, and the N to NV$^-$ conversion yield (typically <10%) [34]. Recently, progress was made to increase the production yield by more than 30% by optimizing the process (e.g., the electron irradiation dose) [35]. However, >70% of the implanted/doped nitrogen atoms are present in the substrate as substitutional nitrogen, known as P$_1$ paramagnetic centers [36,37] or neutral nitrogen-vacancy (NV$^0$) centers [1]. These excesses of nitrogen unavoidably lead to decreased NV$^-$ center spin coherence properties by orders of magnitude [36–38], thus degrading the sensitivity [25]. Also, a large implantation or electron irradiation dose creates undesired defects and local graphitization of the diamond crystal, which could be beyond repair via the thermal annealing process [39,40]. These unwanted defects decrease spin coherence times $T_2^*$ and $T_2$ that impact the overall sensitivity. Therefore, developing low-cost and highly efficient methods in creating high-density NV$^-$ centers from low-concentration nitrogen in the diamond crystal is essential to increase the number of NV$^-$ centers for many scientific and commercial applications such as the fabrication of brighter nanodiamonds for biomedical applications [12,22,33,41].

This work focuses on a quick and cost-effective method of creating high-density NV$^-$ centers in type-IIa chemical vapor deposition (CVD)-grown diamond substrates with as-grown nitrogen concentration of 1 ppm by using high-energy photons from Ar$^+$ plasma. Based on optically detected magnetic resonance (ODMR) and fluorescence confocal microscopy measurements, we estimate the density of newly created NV$^-$ ~ 0.57 ppm (> 50% of conversion yield), distributed homogeneously over 200 µm deep from the diamond surface facing the plasma source. The created NV$^-$ centers exhibit a spin-lattice relaxation time ($T_1$) of 5±0.2 ms and a spin-spin coherence time ($T_2$) of 4±0.5 µs. We estimate a shot noise-limited DC magnetic field sensitivity of ~104 nT Hz$^{-1/2}$ an AC magnetic field sensitivity of ~ 0.12 pT Hz$^{-1/2}$ respectively over an active sample volume of 0.2 µm$^3$, and demonstrate a real-time AC magnetic field sensing with a frequency of up to 90 Hz using the same active sample volume.

### III. Experimental methods and discussion

The NV center is a solid-state spin sensor in a diamond crystal formed by substituting a carbon atom with a nitrogen atom and a vacancy adjacent to it [1] (inset of **Fig. 1b**). This type of nitrogen-vacancy center is known as the neutral NV$^0$ center. If the neutral NV center captures an extra electron, it forms a negatively charged NV$^-$ center that can be photoionized to NV$^0$ with laser or voltage excitations [42,43]. The NV$^-$ center, a spin-1 system, is used for emerging quantum applications, including quantum sensing [9]. The diamond substrates used for this work are single crystal CVD-grown type-IIa diamonds (element six part# 145-500-0055) doped with ~ 1ppm (1.76 $\times 10^{17}$ cm$^{-3}$) of nitrogen (N) throughout the substrate during the growth process. The substrates (labeled 1 and 2) were first cleaned in a tri-acid mixture (1 H$_2$SO$_4$: 1 HNO$_3$: 1 HCLO$_4$) [44] for two hours at 200 °C to remove polishing-related graphitization, followed by rinsing with deionized (DI) water or Isopropyl Alcohol (IPA) and finally drying using compressed nitrogen gas. The dry substrates were then exposed to Argon (Ar$^+$) plasma by using Trion Minilock-Phantom III Inductively Coupled Plasma (ICP) Reactive Ion Etching (RIE) system for 30 s (**Fig. 1(a)**). The plasma process was performed with ICP power at 200 W, RIE power at 50 W, Ar gas flow at 5 sccm, and the ICP-RIE chamber pressure at 10 mTorr. After the plasma exposure, the diamonds were cleaned with the tri-acid mixture for two hours at 200 $^o$C. The dry substrates were then annealed under vacuum (10$^{-8}$ Torr) at 1100 $^o$C for two hours with another subsequent triacid cleaning [31,41]. The clean and dry substrates were then used for optical and spin characterization. We discuss below the mechanisms of NV$^-$ creation using high-energy photons from Ar$^+$ plasma.

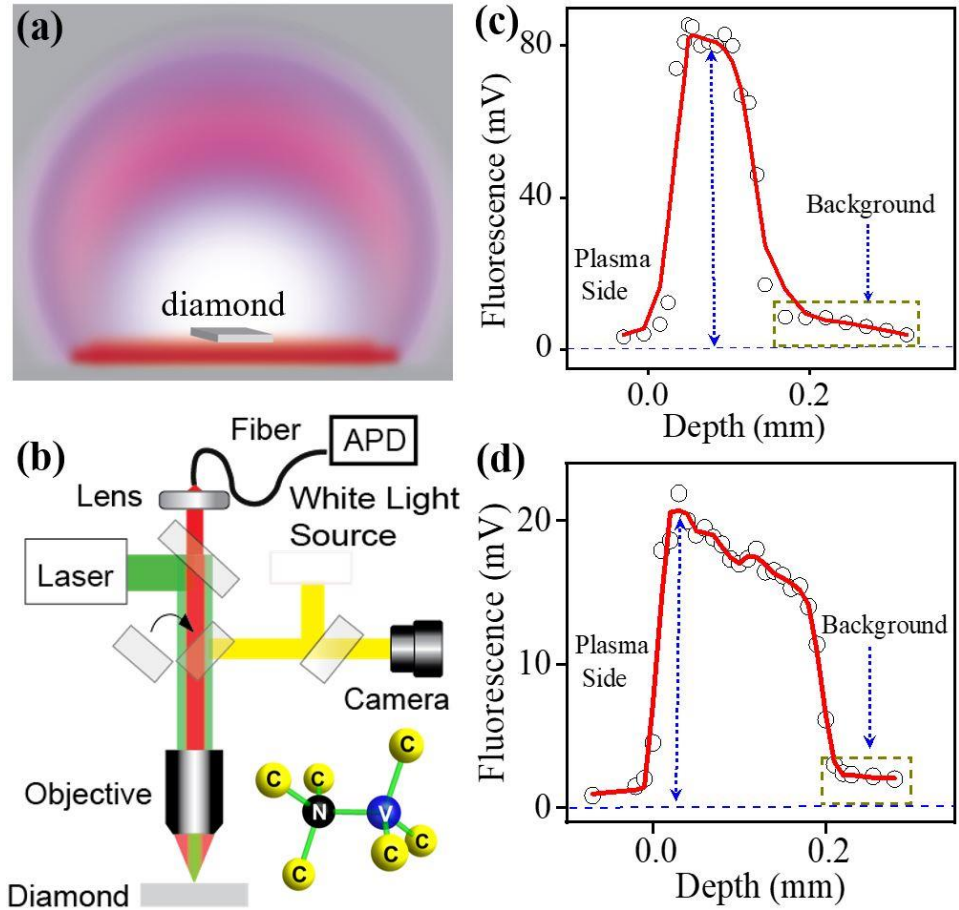

**Figure 1.** **(a)** Representation of the process for making a high-density NV$^-$ layer using Ar$^+$ plasma. **(b)** schematic of the custom-made confocal optical microscope used to characterize the high-density NV$^-$ centers. Inset of **(b)** an NV center in a diamond lattice. The distribution of the created NV$^-$ centers as a function of the depth from the side facing the plasma measured on 0.25 mm (**c**, substrate 1) and 0.5 mm (**d**, substrate 2) thick diamonds prepared with the same conditions.

The NV$^-$ center is a deep bandgap defect located near the mid-gap within the 5.6 eV bandgap of the diamond crystal [1]. NV$^-$ formation needs three processes: substitutional nitrogen atom, adjutant vacancy, and electron capture. There are two well-established NV$^-$ creation methods: *(1)* implantation of nitrogen ($^{14}$N or $^{15}$N) ions followed by high-temperature annealing [45], and *(2)* doping the diamond crystal with N during the growth process and create vacancies by helium ion (He$^+$) implantation [32,33] or electron irradiation [34,46] followed by subsequent high-temperature annealing. In both methods, the implantation/irradiation of N/He ions creates a high density of vacancies due to the broken bonds/dangling bonds. During the high-temperature annealing process, these vacancies position themselves next to substitutional nitrogen atoms forming the $NV^0$ center [1]. When $NV^0$ captures an extra electron, they become $NV^-$. Since the diamond has a large bandgap and NV centers are located near the middle of the bandgap, the thermal energy is insufficient to excite electrons from the valence band to the conduction band to be captured. These difficulties lead to poor yield in N-to-$NV^-$ formation. However, electron irradiation creates broken bonds along with extra electrons, improving $NV^-$ formation efficiency [34,45].

High energy and above bandgap photons generated from the Ar+ plasma were used to create dangling bonds in $SiO_2$ [47–49]. The energy of $Ar^+$ in the plasma is very low, and the depth of the created $NV^-$ center is more than 200 µm from the surface facing the plasma. Therefore, it is conclusive that the enhancement of $NV^-$ formation is primarily due to high-energy photons from $Ar^+$ plasma [50–52]. We hypothesize that the high energy and above bandgap photons from $Ar^+$ plasma could have two effects: (*i*) Direct absorption of UV photons inducing vacancies such as broken bonds/dangling bonds [53]. (*ii*) When the diamond absorbs a UV photon, it creates a shower of electrons (because the photon energy is much larger than the diamond band gap) [54]. These electrons could travel through the conduction band of the diamond and recapture. Therefore, we assume that the high-energy photons would have similar effects as electron irradiation, creating vacancies and supplying electrons for recapturing. The formation of dangling bonds will create more NV centers from doped $N^0/P_1$ centers, similar to electron irradiation [34,46]. The excited extra electrons in the conduction band could be captured to form $NV^-$ centers [54].

*Optical Characterization.*
The fluorescence properties of the diamond substrates were investigated using a custom-built confocal fluorescence microscope [14,55] (**Fig. 1b**) consisting of a green laser (532 nm) for optical excitation of $NV^-$ centers, a permanent magnet (up to 100 mT) to apply a magnetic field for ODMR measurements, a 100x microscope objective with a numerical aperture of 0.8 NA to focus light on the diamond substrate, and fluorescence collection optics including 650 nm edge pass filter, focusing lens, 9/125 single-mode fiber, single photon counter modules (SPCM) used for confocal imaging and spin measurements in **Fig. 2** and **Fig. 3**, and avalanche photodetector (APD) used for measurements in **Fig. 1** and **Fig. 4**.

The diamond substrate is excited with a green (532 nm) laser, and the fluorescence is measured at different depths from the surface facing the plasma, **Figs. 1c**, **1d**. We repeat the process to several locations and different substrates thicknesses. The measurement results from two of the representative CVD diamond substrates of thicknesses, 0.25 mm (substrate 1) and 0.5 mm (substrate 2) are shown in **Fig. 1c** and **Fig. 1d**. at a green laser power of 20 mW and 5 mW, respectively. The fluorescence intensity in both diamonds is an order of magnitude higher for the surface facing the Ar+ plasma than the other side, which has only background fluorescence coming from the as-grown $NV^-$ centers (< 0.1 ppm). The newly created $NV^-$s are uniformly distributed over a depth of 150-200 µm depending on the substrate thickness, as shown in **Fig. 1c** and **Fig. 1d**.

To estimate the spatial distribution of the $Ar^+$ plasma-created $NV^-$ centers, we performed fluorescence imaging on the surfaces of substrate 2 facing plasma (**Fig. 2a**) at a green laser power of 0.5 mW. **Fig. 2b** shows the line cut profile (dashed line) taken on **Fig. 2a**. It is conclusive that the Ar+ plasma created $NV^-$ centers are uniformly distributed. We compare the fluorescence intensity from the newly created ensemble $NV^-$ centers in substrate 1 with the one collected from an electronic grade (EL) diamond with single NV$^-$ centers (created by ion beam implantation of $^{15}$N ions at a dose of $5\times10^9$ cm$^{-2}$ and energy of 30 keV) by using the confocal microscope under similar experimental conditions (at saturation). Since our SPCM saturates at a count rate of 10 Mc/s corresponding to a laser power of 3 mW, we used an APD to detect the fluorescence of the plasma-created NV$^-$s. In **Fig. 2c,** we plot the fluorescence of substrate 1 as a function of green laser power by converting the APD detected voltage (mV) to counts per second (c/s) and found a photon count rate of 400 Mc/s at saturation (> 30 mW). Based on single NV$^-$ measurements (maximum count rate of 0.02 Mc/s at saturation) on the EL implanted diamond (inset of Fig. **2d**), we estimate that the side facing Ar$^+$ in substrate 1 contains > 20,000 NV$^-$ centers over the active volume of 0.2 µm$^3$, which translates to an NV$^-$ density of ~$10^{17}$ cm$^{-3}$ (= 0.57 ppm). To confirm that the detected fluorescence comes from NV$^-$ centers, we performed fluorescence vs wavelength measurements on the side of substrate 2 facing Ar$^+$ plasma by using spectrometer TRIAX 320 (Horiba), **Fig. 2d**. The observed spectrum consists of a typical NV$^-$ curve with high fluorescence in the wavelength range of 650- 760 nm with zero phonon line (ZPL) peaks for NV$^0$ and NV$^-$ at 574 and 637, nm respectively [1].

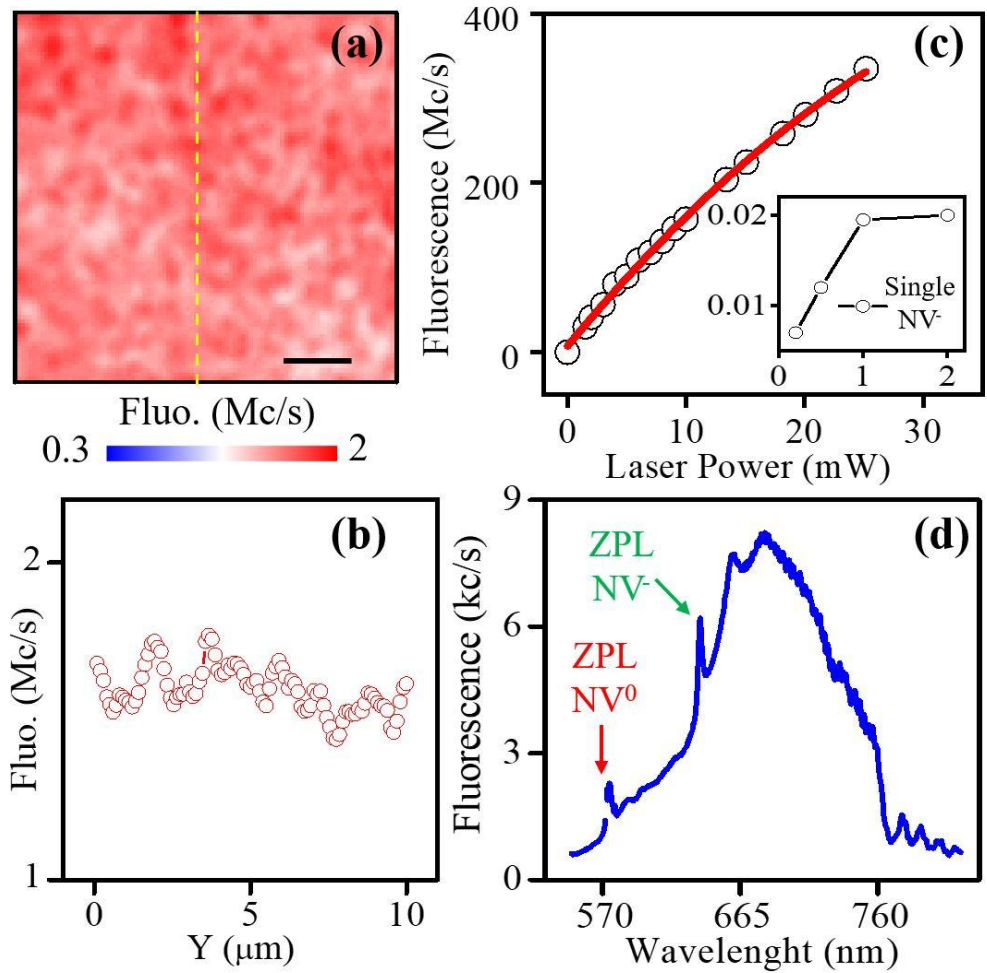

**Figure 2.** *Optical characterization of the Ar$^+$ plasma photon irradiated diamond substrates.* Fluorescence image of the surface facing the Ar$^+$ plasma (**a**). (**b**) Vertical line cut of fluorescence spatial profile taken across the dashed lines in (**a**). (**c**) The fluorescence of substrate 1 was detected by APD as a function of green laser power. Inset of (**c**) fluorescence intensity of EL diamond with single NV$^-$ centers as a function

of green laser power (20 Kc/s at a laser power of 2 mW). **(d)** Fluorescence spectrum acquired on the surface facing the plasma of substrate 2. Scale bar in **(a)** is 2 μm.

*Spin-characterization.*

The $NV^-$ center-based sensing application requires narrow resonance linewidth and sufficiently long coherence and relaxation time. The spin-resonance properties of the $Ar^+$ plasma created high-density $NV^-$ are scrutinized using well-established electron spin resonance measurement methods [14,15,56]. **Fig. 3** shows the outcome of the measurements on substrate 1 performed using the custom-built confocal microscope with SPCM modules and at an applied magnetic field of 8.6 mT applied along [111] orientation of the (100) diamond and at laser power of 1 mW. We discuss the ODMR measurements in detail below.

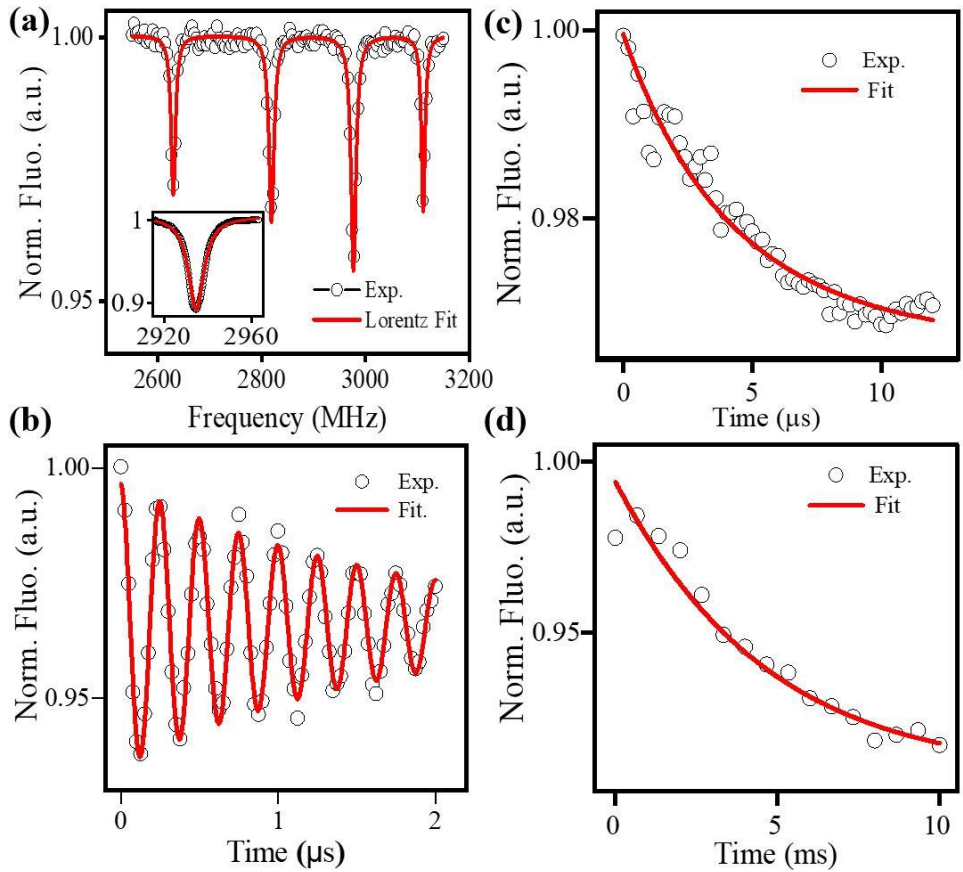

**Figure 3**. **(a)** CW-ODMR spectrum measured (scattered open circles) at an applied field of 8.6 mT. The solid line curve in **(a)** is Lorentzian fits for the NV ODMR peaks. Insert in **(a)** is the cw-ODMR spectra acquired at 2.3 mT field with optimized laser and MW parameters. **(b)** Rabi-nutation measurement which shows the fluorescence intensity (scattered open circles) vs the duration of applied MW pulse at MW frequency of 2629 MHz and a magnetic field of 8.6 mT. The solid line curve is a fit to the experimental data (see the main text). **(c)** Normalized measured (scattered open circles) fluorescence intensity vs echo time between p pulses fitted (solid line) with an exponential decay function with a decay = $T_2$. **(d)** Normalized measured (scattered open circles) fluorescence intensity vs time, fitted with an exponential decay function with a decay = $T_1$.

**A. cw-ODMR:** The ODMR spectrum of $NV^-$ centers is measured by alternating the microwave (MW) power between OFF and ON states. In the OFF state, $NV^-$ centers are continuously (10-100 ms pulses) pumped into the bright |0⟩ state, while in the ON state, fluorescence is reduced as spins are driven into |∓1⟩ states through the intersystem crossing to metastable singlet states [1]. The normalized fluorescence intensity is recorded as a function of the MW frequency at the applied magnetic field oriented along [111] direction of the (100) diamond substrate (**Fig. 3a**). The resonance full-width-at-half-maximum linewidth $\Gamma$ is 8.6 MHz, similar to the ensemble NV⁻ centers created by ¹⁴N substitution [36]. The applied magnetic field breaks the degeneracy of the |∓1⟩ state and leads to a pair of transitions for each NV⁻ orientation whose frequencies depend on the field projection along the NV symmetry axis. There are four sub-ensembles of NV centers with different symmetry axes; thus, a full ODMR spectrum contains 8 peaks (4 for |0↔1⟩ transition and 4 for |0↔−1⟩ transition) [57]. To estimate the DC magnetic field sensitivity, we acquired the high-field (|0↔1⟩) cw-ODMR spectrum at an applied field of 2.3 mT with optimized measurement parameters such as the laser power and MW power (inset of **Fig. 3a**) [45]. The minimum detected DC magnetic field within the shot noise is given by [14,25,58]: $\eta_{CW} \cong 4\Gamma (3\sqrt{3}\gamma_{NV} C)^{-1} (R)^{-1/2}$, where $\gamma_{NV} = $ 28 GHz/T is the gyromagnetic ratio of the electron spin, $C$ is the ODMR peak contrast, $R$ is the NV photon-detection rate. By using the parameters of the measurements (inset of **Fig. 3a**, $R$ = 200 mV = 400 M counts/s measured by using Thorlabs APD (APD440A) at laser power of 20 mW, $\Gamma$ = 8.6 MHz, C = 0.1123) we estimate a DC sensitivity of ~104 nT/Hz$^{1/2}$ over 0.2 µm³ active volume.

B. **Rabi Nutation:** We performed Rabi nutation measurements to check the $T_{2,Rabi}$ decay [45] and know the π pulse length required to measure NV⁻ spin coherence lifetimes $T_2$ and $T_1$ of the plasma created NV⁻ centers (see below). We applied an MW frequency of 2629 MHz at the |0↔−1⟩ peak along [111] orientation (the left ODMR peak in **Fig. 3a**) and recorded the NV fluorescence intensity vs the MW duration time $t$. **Fig. 3b** displays the measured Rabi oscillations fitted with a function $\sin(\omega_{Rabi} t) \exp(-t/T_{2,Rabi})$ ($\omega_{Rabi}$ is the Rabi frequency ~ 4 MHz). The π pulse is 124 ns and $T_{2,Rabi}$ decay is 1.74 µs that can be extended to > 50 µs upon a periodic reversal of the π phase [59].

C. **Transverse spin relaxation time ($T_2$):** Sensing weak dynamic (magnetic, electrical, or thermal) signals requires both high density of NV⁻ centers and longer $T_2$ relaxation times. We used a standard three-pulse Hahn-echo protocol to measure the $T_2$ of the Ar⁺ plasma created NV⁻ centers [11,14,15]. The pulse sequence consists of a $\frac{\pi}{2} - \tau - \pi - \tau - \frac{\pi}{2}$ applied on the OMDR resonance peak at 2629 MHz in **Fig. 3a**, and the integrated NV fluorescence intensity is recorded as a function of the total evolution time $2\tau$ ($\tau$ is the time between π/2 and π pulses). **Fig. 3c** shows a fast exponential decay of the Hahn-echo envelope (4 µs), explained by the strong dipolar interactions between the NV⁻s and N paramagnetic centers (1 ppm in our CVD diamond). The contribution of ¹³C spins in the NV echo signal at an applied magnetic field of 8.6 mT is negligible since the signal decays before the first ¹³C revival (20 µs). The estimated sensitivity for dynamic (AC) magnetic fields is [25]: $\eta_{AC} \cong (\gamma_{NV} D)^{-1} 1/(R T_{meas})^{1/2} \exp(T/T_2)$. $T$ is the full field interrogation time, $D$ is the spin echo contrast, and $T_{meas}$ is the measurement averaging time. By using the parameters of our measurements ($T_2$ = 4 µs, $R$ = 200 mV = 400 M counts/s, $D$ = 0.02, $T_{meas}$ = 1 s, $T$ = 2 µs) the sensitivity to AC magnetic fields is 0.12 pT Hz$^{-1/2}$. By using dynamical decoupling pulse sequence [25,31,41,60] $T_2$ can be extended to > 100 µs for NV⁻ ensemble spins, and the AC sensitivity can be pushed to a few fT Hz$^{-1/2}$ [61].

**D. Longitudinal spin relaxation time ($T_1$)**: We measured the longitudinal spin relaxation time ($T_1$) of the Ar$^+$ plasma-created NV$^-$ centers using standard $T_1$ measurements [15,62]. We applied a sequence consisting of two laser pulses for NV initialization/readout (10 μs & 0.3 μs in duration) with and without π pulse and recorded the subtracted fluorescence intensity as a function of the time $t$ between the initialization and readout pulses. We plot the result of the $T_1$ measurements in **Fig. 3d** and found a $T_1 \sim 5$ ms, similar to the values measured in CVD and EL diamonds. $T_1$ measurements are useful when using spin-correlation pulse protocols to bypass $T_2$ and improve AC sensitivity [15,31].

## IV. Applications

High densities of NV$^-$s are favored for precision magnetometry due to improved signal-to-noise and sensitivity from $1/\sqrt{N}$ enhancement [25]. High densities also provide high fluorescence intensity, which can be measured by regular cheap photodetectors reducing the complexity of using SPCMs. To demonstrate magnetic field sensing applicability using the Ar$^+$ plasma-created NV$^-$ centers, we implemented a magnetic field tracking method described by Welter et *al.* [17]. A 0.15 mT oscillatory magnetic field of various frequencies is applied to the NV$^-$ sensors, which track the changes in the applied magnetic field in real-time, as shown in **Fig. 4**. We connected the tracking output voltage with an oscilloscope and recorded the time transient of the tracking signal along with applied magnetic field signals. As shown in **Fig. 4a**, the applied magnetic field changes 0.3 mT (peak-to-peak) over 20 ms time, and the tracking system can follow those changes in real-time. **Fig. 4b** shows the fast Fourier transform (FFT) of the measured tracking signals for all the applied oscillatory magnetic fields.

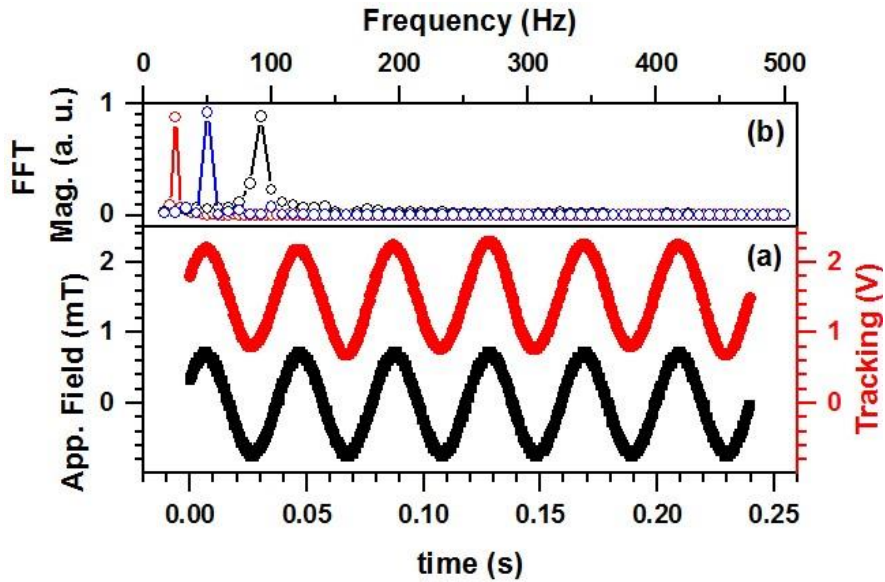

**Figure 4.** Real-time magnetic field measured using the NV$^-$ centers created by high-energy photons from Ar$^+$ plasma over an active volume of 0.2 μm$^3$.

## V. Conclusion

In summary, we demonstrated a low-cost and highly efficient method to create high-density NV-centers on CVD-diamond crystal containing as-grown nitrogen concentrations of 1 ppm. The measurements show that the high-density NV⁻ layer is distributed over 150-200 µm from the surface facing the photons created by Ar⁺ plasma. It demonstrates that the described process creates an NV⁻ density of 0.57 ppm, yielding more than 50% N-to-NV⁻ conversion. We measured a DC magnetic field sensitivity of ~ 104 nT Hz$^{-1/2}$ and an AC magnetic sensitivity of ~ 0.12 pT Hz$^{-1/2}$ measured over 0.2 µm³ active sample volume. The measured relaxation properties are $T_1 = 5$ ms and $T_2 = 4$ µs. We found that the fluorescence intensity from the NV⁻ centers within 0.2 µm³ volume is detectable using regular photodetectors (e.g., APD), which is beneficial for on-chip and commercial adaptation [27]. The measurements show that using the Ar⁺ plasma-created high-density NV⁻ can be used to measure magnetic fields quickly and efficiently at a rate over 10 mT/s with an active sample volume of 0.2 µm³. Our methodology could find applications in magnetometry where thick NV layers (150-200 um) are needed to measure magnetic fields generated from big cells labeled with magnetic nanoparticles [63] or planetary bodies for paleomagnetic analysis of complex rocks such as meteorites that have heterogeneous magnetizations ≤ 200 µm [64].

## VI. Acknowledgments

K.A. would like to acknowledge the support of the National Science Foundation/EPSCoR RII Track-4 Award OIA-2033210. The research was partially performed in the Nebraska Nanoscale Facility, which is supported by the National Science Foundation under Award ECCS: 2025298 and the Nebraska Research Initiative. P.B.K. would like to acknowledge the support of the Wichita State University Convergence Science Initiative Program. A. L. would like to acknowledge the support of the National Science Foundation/EPSCoR RII Track-1: Emergent Quantum Materials and Technologies (EQUATE), Award OIA-2044049.

## VII. AUTHOR DECLARATIONS

### a. Conflict of Interest

The authors have no conflicts to disclose.

### b. Author Contributions

K.A. conceived the experiment and supervised the overall study. P.B.K., R.T., M.D., A.L., and K.A. performed the measurements and analyzed the data. A. E. A. helped in data analysis and figure preparation for the manuscript. A.L. and K.A. wrote the manuscript with help from all authors. P.B.K. and R.T. contribute equally.

## VIII. DATA AVAILABILITY

The data supporting this study's findings are available from the corresponding author upon reasonable request.